\documentclass[twocolumn,showpacs,aps,prl,superscriptaddress,floatfix]{revtex4}
\usepackage{graphicx}
\usepackage{dcolumn}
\usepackage{bm}

\begin{document}

\preprint{}

\title{Susceptibility anisotropy in an iron arsenide superconductor revealed by x-ray diffraction in pulsed magnetic fields.}

%
\author{J.P.C. Ruff}
\email{jpcruff@aps.anl.gov}
\affiliation{The Advanced Photon Source, Argonne National Laboratory, Argonne, Illinois 60439, USA}
\author{J.-H. Chu}
\affiliation{Geballe Laboratory for Advanced Materials, Stanford University, Stanford, CA 94304, U.S.A.}
\affiliation{Department of Applied Physics, Stanford University, Stanford, CA 94304, U.S.A.}
\author{H.-H. Kuo}
\affiliation{Geballe Laboratory for Advanced Materials, Stanford University, Stanford, CA 94304, U.S.A.}
\affiliation{Department of Materials Science and Engineering, Stanford University, Stanford, CA 94304, U.S.A.}
\author{R.K. Das}
\affiliation{The Advanced Photon Source, Argonne National Laboratory, Argonne, Illinois 60439, USA}
\author{H. Nojiri}
\affiliation{Institute for Materials Research, Tohoku University, Katahira, Sendai 980-8577, Japan}
\author{I.R. Fisher}
\affiliation{Geballe Laboratory for Advanced Materials, Stanford University, Stanford, CA 94304, U.S.A.}
\affiliation{Department of Applied Physics, Stanford University, Stanford, CA 94304, U.S.A.}
\author{Z. Islam}
\affiliation{The Advanced Photon Source, Argonne National Laboratory, Argonne, Illinois 60439, USA}

\begin{abstract} 

In addition to unconventional high-T$_{\bf{c}}$ superconductivity, the iron arsenides exhibit strong magnetoelastic coupling and a notable electronic anisotropy within the \emph{a-b} plane.  We relate these properties by studying underdoped Ba(Fe$_{\bf{1-x}}$Co$_{\bf{x}}$)$_{\bf{2}}$As$_{\bf{2}}$ by x-ray diffraction in pulsed magnetic fields up to 27.5 Tesla.  We exploit magnetic detwinning effects to demonstrate anisotropy in the in-plane susceptibility, which develops at the structural phase transition despite the absence of magnetic order.  The degree of detwinning increases smoothly with decreasing temperature, and a single-domain condition is realized over a range of field and temperature.  At low temperatures we observe an activated behavior, with a large hysteretic remnant effect.  Detwinning was not observed within the superconducting phase for accessible magnetic fields.

\end{abstract} 

\maketitle 

Both the copper-oxide and the iron-arsenide families of high-T$_c$ superconductors present layered structures of strongly correlated magnetic planes\cite{mrnorman}.  That the electronic interactions within these planes show a significant anisotropy in both families is becoming increasingly clear.  Various bulk and local probes infer an underlying ``electronic nematicity'' in the phase diagrams of these compounds, generally proximate to the superconducting phase\cite{mrnorman,jhchu_science,davis_science1,davis_science2,lavrov_magdetwin,jhchu_magdetwin,kivelson,sachdev,dai_nematic,xu_nematic,sefat_nematic,kim_nematic,clamp_arpes}.  In the case of underdoped Ba(Fe$_{1-x}$Co$_x$)$_2$As$_2$, it has been proposed that this collective electronic tendency to break rotational symmetry is the driving force for the tetragonal-to-orthorhombic structural phase transition, the suppression of which by doping leads to high temperature superconductivity\cite{xu_nematic,sefat_nematic,kim_nematic,clamp_arpes}.  For proposed spin-nematic or orbital-ordering scenarios\cite{kivelson,kim_nematic}, a significant magnetic anisotropy may be expected to develop coincidently with the othorhombic distortion.  The observation of this anisotropy is therefore of interest.

Unfortunately, experiments seeking to uncover nematicity can be hampered by transformational twinning effects, which occur at the tetragonal-to-orthorhombic phase transition and result in roughly equal populations of orthorhombic domains which are rotated by roughly 90$^o$ with respect to each other\cite{jhchu_science,lavrov_magdetwin,jhchu_magdetwin,goldman}.  In a twinned orthorhombic crystal, it is practically impossible to resolve in-plane anisotropy in properties such as resistivity or susceptibility, due to the averaging over twin domains.  Nevertheless, experimental progress has been made by either probing length scales much smaller than the twin domain size\cite{davis_science2}, or by ``detwinning'' crystals through the application of uniaxial stress\cite{jhchu_science,clamp_arpes,kim_nematic}.  These investigations have unanimously returned evidence for an electronic inequivalence between the orthorhombic $a$ and $b$ directions which exceeds what would be expected from the small difference in lattice constants.  However, critics may conject that the application of uniaxial stress is itself a symmetry-breaking perturbation which can induce extra anisotropy.  Although this seems not to be the case experimentally\cite{jhchu_piezo}, it is nevertheless desireable on aesthetic grounds to probe the anisotropy in the absence of secondary symmetry breaking via stress.

Here we report a ``contact-free'' experimental probe of in-plane anisotropy in the iron arsenide superconductors, which can be applied over a range of temperatures.  Our approach exploits the anisotropy of the magnetic susceptibility to detwin crystals through the application of magnetic field, rather than uniaxial pressure.  Our measurements are predicated on the assumption that $\chi_b \neq \chi_a$, as suggested in Ref.\cite{jhchu_magdetwin}.  Magnetic fields can be applied within the $a$-$b$ plane such that the field aligns simultaneously along the $a$-axis for one species of twin, and the $b$-axis for the other species.  The magnetic detwinning effect is then understood to arise from the difference in the free energy between the two twin populations in applied magnetic field, which exerts an effective force on the twin boundaries.  Magnetic detwinning has been demonstrated previously for both the copper-oxide superconductors\cite{lavrov_magdetwin} and the iron-arsenides\cite{jhchu_magdetwin} by transport and optical measurements, however, our approach uniquely allows a direct quantitative measurement of the variations in twin volume fractions over a wide range of temperature and applied magnetic field.  Using newly developed time-resolved x-ray diffraction techniques in pulsed magnetic field\cite{zahir_split,zahir_strip,ruff_pmag_tto}, we uncover the full B-T phase diagram for magnetic detwinning in an single crystal of Ba(Fe$_{1-x}$Co$_x$)$_2$As$_2$ with $x=0.045$.  From these measurements we infer the temperature evolution of the susceptibility anisotropy on cooling through the four characteristic phases of the underdoped iron arsenides (tetragonal paramagnet, orthorhombic paramagnet, orthorhombic spin density wave, and superconductor).

Ba(Fe$_{1-x}$Co$_x$)$_2$As$_2$ with $x=0.045$ exhibits three well-defined phase transitions, with the structural transition occurring at 70 K, the spin density wave transition occurring at 60 K, and the superconducting transition occurring at 13 K\cite{jhchu_phasediag}.  These transitions  were confirmed by resistivity measurements of our crystal.  We prepared our sample as a thin square plate with edges cut parallel with the tetragonal $a$-$b$ axes, and mounted it at the bottom of a sapphire post.  The crystal was attached to the sapphire using GE varnish at the top corner only, such that the majority of the sample was freely hanging, to minimize any clamping stress. The sapphire rod and sample were subsequently inserted into the bore of a split-pair pulsed magnet, designed to facilitate x-ray diffraction in magnetic fields as high as $\sim 30$ Tesla.   Operation of this instrument at the Advanced Photon Source (APS) is detailed elsewhere\cite{zahir_split,zahir_strip,ruff_pmag_tto}.  Data were collected at beamline 6ID-B, using 16.2keV photons to illuminate a (0.2 mm$\times$0.4 mm) area near the bottom of the sample.  Results obtained in the absence of magnetic fields are illustrated in Figure 1, where the tetragonal-to-orthorhombic transition at T$_S$ = 70 K is readily confirmed.  The tetragonal (2,2,0) Bragg peak splits into the orthorhombic (4,0,0) and (0,4,0) at the transition, with the two peaks resulting from the two types of twin domains.  Retaining tetragonal notation, the observed splitting along the (H,0,0) direction into peaks of the approximate form $($($2\pm\Delta$),2,0$)$ identifies the orientation of the twin planes within the illuminated volume of the sample\cite{note1}.  Splitting along (H,0,0) corresponds to twin planes stacked perpendicular to (0,K,0), as illustrated in Figure 1(d) and described in Ref.\cite{goldman}.  In principle, we could also have observed a second set of peaks split in (0,K,0), which would result from symmetry-related twinning planes perpendicular to (H,0,0).  However, the volume probed in these experiments does not manifest those peaks, and therefore lacks those rotated twin planes.  

\begin{figure} 
\centering 
\includegraphics[width=9.3 cm]{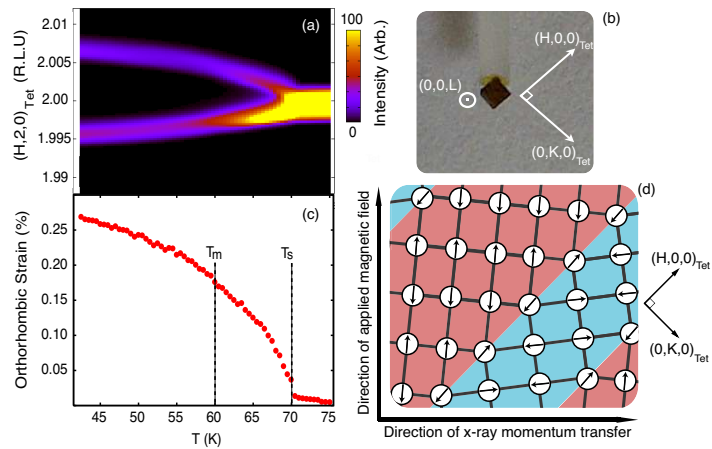}
\caption {(a) High resolution x-ray diffraction shows the (2,2,0) Bragg peak splitting at the structural phase transition (T$_S$ $\sim 70$ K).  (c) Orthorhombic strain, defined as $2 (a-b)/(a+b)$, as a function of temperature showing the relative difference in the lattice constants below T$_S$.  The spin density wave transition T$_m$ is also indicated for reference.  (b)  Our $x=0.045$ single crystal, mounted on a sapphire post.  We perform diffraction in the horizontal [HHL] plane, in transmission through the c-face of our crystal.  We apply magnetic fields along the vertical $[1\bar10]$ direction, coaxial with the sapphire post.   (d) Schematic of twin domains within the $a$-$b$ plane.  Orthorhombic strain and the density of twinning is exaggerated for effect.  The highly anisotropic magnetic ordering pattern of Fe spins is also illustrated.}
\label{fig:1}
\end{figure}

\begin{figure} 
\centering 
\includegraphics[width=8.5 cm]{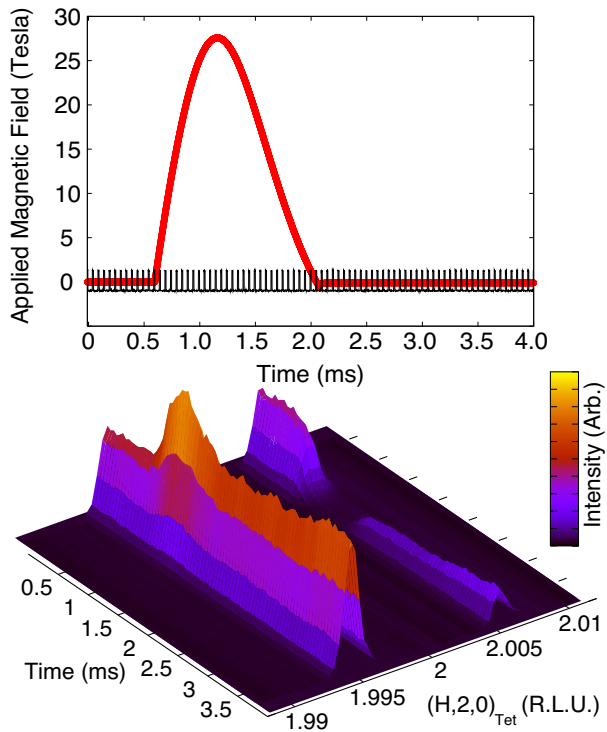}
\caption {Evolution of twin reflections at T = 30 K, while magnetic field is pulsed to 27.5 Tesla.  Top:  The applied magnetic field as a function of time.  The square pulses denote the 20 kHz framerate collection of our one-dimensional strip detector, which generates a series of 30-$\mu$s exposures of the Bragg scattering along arcs in reciprocal space.  Bottom:  Arcs collected at both twin peak positions and projected onto the tetragonal (H,2,0) line in reciprocal space.  The peak at higher momentum transfer is completely suppressed at high field, with the intensity transferred to the other peak.  This indicates a complete elimination of the twin domains with magnetic field applied along the longer $a$-axis, in favor of the domains with magnetic field along the shorter $b$-axis.}
\label{fig:2}
\end{figure}

\begin{figure*} 
\centering 
\includegraphics[width=15 cm]{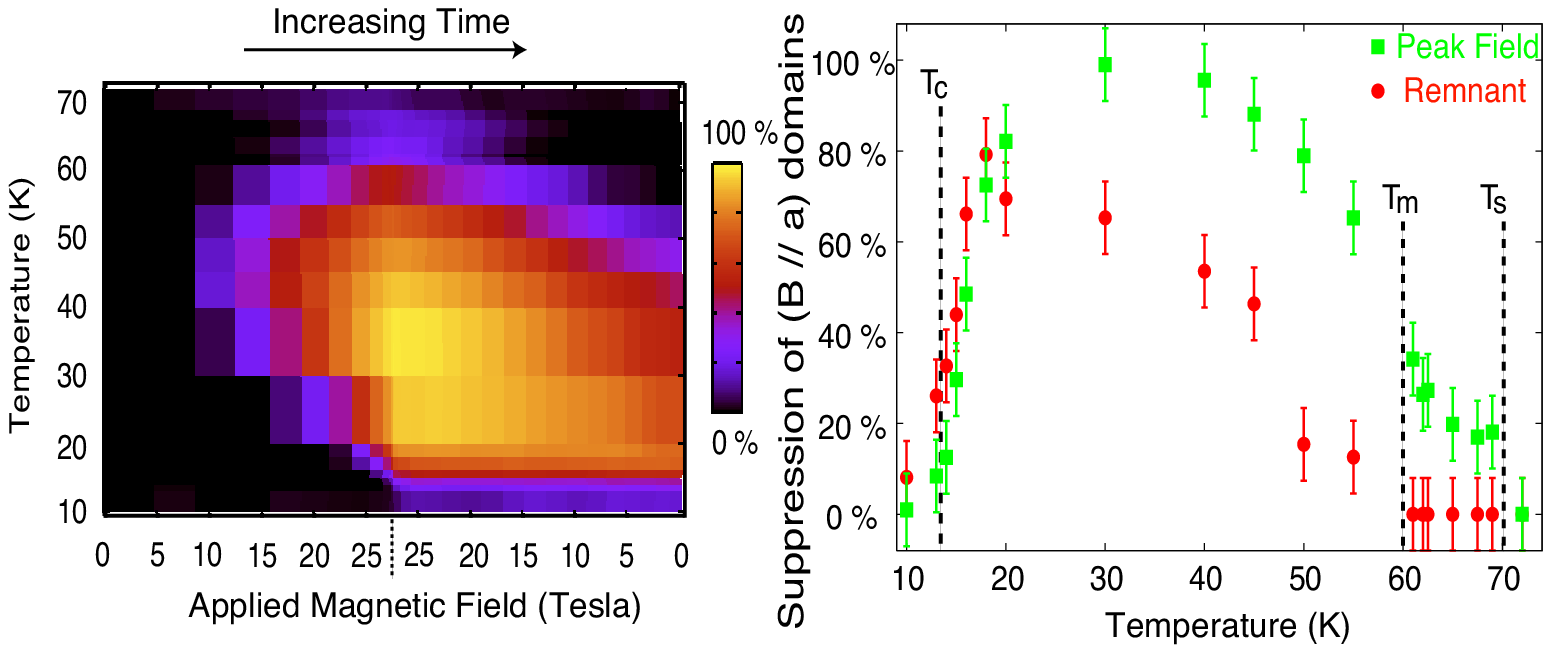}
\caption {Suppression of the B $\parallel a$ twin domains, as a function of temperature and applied magnetic field.  Here, 100$\%$ suppression signifies a completely detwinned volume, while 0$\%$ suppression corresponds to a twin population balance that is completely unchanged by applied magnetic fields.  Left:  Data collected on the $($($2+\Delta$),2,0$)$ Bragg feature, after cooling in zero field from 80 K and pulsing magnetic fields to 27.5 Tesla.  Data for rising and falling fields are shown, with time during the pulse increasing along the x-axis.  Right:  Cuts through the same data, highlighting the effect at the highest measured peak fields, and also the residual remnant detwinning after the pulse is over.  The temperatures of the structural (T$_s$), magnetic (T$_m$), and superconducting (T$_c$) phase transitions are indicated.  Detwinning in the paramagnetic phase is clear.}
\label{fig:3}
\end{figure*}

A schematic depiction of our pulsed-magnetic-field measurement is shown in Figure 1(d).  We apply magnetic fields along the shared $a$/$b$ direction of the two twin populations, and measure the Bragg scattering from each domain independently as a function of time during the pulse.  The Bragg peak which appears at higher momentum transfer results from twins oriented with the shorter ($b$-axis) lattice constant in the scattering plane, and therefore the longer ($a$-axis) lattice constant along the magnetic field direction.  The converse is true for the Bragg peak which appears at lower momentum transfer.  For the proposed susceptibility anisotropy where $\chi_b > \chi_a$, we expect the twin domains with the $b$-axis aligned along the magnetic field to grow, at the expense of the domains with the $a$-axis along the field direction.  Therefore, x-ray scattering intensity should be depleted for the twin peak at higher momentum transfer $($($2+\Delta$),2,0$)$, and should be enhanced for the twin peak at lower momentum transfer $($($2-\Delta$),2,0$)$.  The results of a measurement at T = 30 K, in pulsed magnetic fields up to 27.5 Tesla, are illustrated in Figure 2.  As the magnetic field is increased as a function of time, both twin peaks respond as expected.  At peak field, we observe a full suppression of the B $\parallel a$ peak, with the scattering intensity transferred to the B $\parallel b$ peak.  This is indicative of a complete detwinning of the illuminated volume.  In reference to the schematic illustration in Figure 1(d), this is equivalent to a single ``blue'' domain in high magnetic fields.  Our conjecture of an anisotropic-susceptibility-based origin of the detwinning is corroborated by the constant peak position during the pulse - the scattering angles and thus the lattice constants associated with the two domains do not change as a function of magnetic field.  Therefore, the effect does not appear to be magnetostrictive.  It is also worth noting that while some scattering intensity is recovered at the $($($2+\Delta$),2,0$)$ position after the pulse, it is clear that an irreversible change that has occurred.  The crystal is residually detwinned after the pulse.

The complete detwinning effect discussed above was observed in the orthorhombic and magnetically ordered phase.  In this region, some anisotropy in the magnetic susceptibility is an expected consequence of the collinear antiferromagnetic order.  This is also the phase in which partial magnetic detwinning has been previously demonstrated to occur\cite{jhchu_magdetwin}.  The prime interest of the present study is to investigate the detailed temperature and magnetic field dependence of the detwinning effect, particularly with respect to the magnetically disordered and superconducting phases.  In order to parameterize a wide region of phase space, we have focused on the suppression of the B $\parallel a$ domains as magnetic fields are pulsed from a zero-field-cooled condition at various temperatures.  We have confirmed the conjugate enhancement effect on the B $\parallel b$ domains at a few representative temperatures only.  The results of these measurements are illustrated in Figure 3.  For temperatures above the structural phase transition, we observe no effect of magnetic field, limiting any magnetostrictive changes in lattice constant to be less than one part in 10$^4$.  However, immediately below the structural transition, we measure an anisotropic response.  We have observed clear detwinning in the temperature region 70 K $>$ T $>$ 60 K, wherein the crystal is orthorhombic but paramagnetic.  This unambiguously demonstrates \emph{an anisotropic magnetic susceptibility that develops coincidently with the structural phase transition}, well in advance of antiferromagnetic order.  This is a central result of this work, consistent with the notion of a spin or orbital driven origin for the orthorhombic distortion.  In the paramagnetic but orthorhombic region, the detwinning effect is reversible, and reaches a maximum population imbalance of $\sim 70/30$ at our peak field of 27.5 Tesla.  As the temperature is lowered, the magnitude of detwinning at all field values increases smoothly, without any obvious anomaly at the magnetic ordering transition.  This trend persists to about T $\sim$ 50 K, below which two significant changes develop.  First, while the detwinning at peak field continues to increase, the detwinning at low field is suppressed, evincing an activated behavior.  Second, the residual remnant detwinning discussed above begins to set in.  We note that magnetic detwinning, as we understand it, requires two ingredients:  anisotropic magnetic susceptibility and the ability of twin boundaries to move.  The absence of detwinning at low temperatures and low magnetic fields therefore implies either a kinetic inhibition or a restoration of susceptibility isotropy.  The simultaneous development of residual detwinning implies the former.  The most compelling argument for the latter would note that the detwinning effect drops to zero at exactly the superconducting transition temperature, as one might expect for a case where electronic anisotropy competes with superconductivity\cite{sachdev_nematic}.  However, this is potentially misleading.  Inspection of Fig. 3 reveals an activation field that increases approximately linearly on cooling, and which therefore crosses our peak field of 27.5 Tesla at T$_c$ potentially by coincidence\cite{note2}.  A final possibility is that the onset field represents a magnetic phase boundary, complete with hysteresis, which must be crossed before the detwinning effect can set in.  Further measurements in higher magnetic fields and at other doping levels promise to shed light on these important outstanding questions.  Meanwhile, the present measurements clearly demonstrate that $\chi_b > \chi_a$ in the orthorhombic phase, both above and below the spin density wave transition temperature, and that this anisotropy is manifest at the temperature scale of the structural phase transition.

We offer a brief further comment on the nature of the residually detwinned state which remains after the magnetic field pulse.  The observed remnant population imbalance between the two types of twin domains, after pulsing to 27.5 Tesla at low temperatures, is comparable to the degree of detwinning that can be achieved through uniaxial strain\cite{jhchu_science}.  We have observed a residual population imbalance that is stable for hours, and possibly longer.  We have also noted a sharper rocking curve for the Bragg reflections in the remnant phase, indicative of an improved crystal quality after defect twin planes are removed through exposure to magnetic fields.  Therefore we conject that treatment with magnetic fields offers a promising alternative to strain for generation of superior quality, detwinned samples of iron arsenide superconductors, provided the samples can be kept cold and the magnetic fields can be applied in-situ prior to measurement.

In this Letter, we have reported an extensive characterization of the magnetic detwinning of an underdoped iron arsenide superconductor, and argued that these measurements can be used to parametrize the temperature dependent susceptibility anisotropy.  Our results reveal yet another manifestation of electronic nematicity in the iron arsenides, which develops above the magnetic ordering transition and can be observed without the use of symmetry-breaking uniaxial strain.  The range of magnetic fields and temperatures which are accessible by pulsed-field diffraction makes our technique uniquely informative in regards to magnetic anisotropy in the iron arsenides, and future investigations of the type reported here will provide an important new perspective throughout the extended phase diagram of these materials.  Future progress will require a more complete understanding of twin domain kinetics in pulsed magnetic fields, which must be modelled accurately in order to extract a more quantitative measure of the susceptibility.  In addition, the details of the coupling of proposed nematic order parameters to magnetic fields also need to be computed.  It is our hope that this letter stimulates new theoretical investigation into these open questions.

Work performed at Stanford was supported by the U.S. Department of Energy (DOE), Office of Basic Energy Sciences, under contract DE-AC02-76SF00515. Use of the APS is supported by the DOE, Office of Science, under contract DE-AC02-06CH11357.  A part of this work was enabled by the International Collaboration Center at the Institute for Materials Research (ICC-IMR) at Tohoku University.  HN acknowledges KAKENHI No. 23224009 from MEXT.  JPCR acknowledges the support of the Argonne Director's Fellowship, and of NSERC of Canada.



\begin{thebibliography}{22}
\bibitem{note1}
Due to the temperature-dependent variation of the average of the orthorhombic $a$ and $b$ lattice constants, which is not fixed at tetragonal $a$, the center of mass of total scattering from the split peaks deviates slightly from the tetragonal (2,2,0) position, and the identification of the two peaks as $($($2\pm{\Delta}$),2,0$)$ is approximate.
\bibitem{note2}
The skin depth is much larger than the sample dimensions.  Estimating an average in-plane resistivity of $5 \mu \Omega \cdot m$ and a pulsed field frequency of 330 Hz results in a prediction of hundreds of mm for flux penetration above Tc, which is orders of magnitude larger than sample dimensions in all directions. The suppression of detwinning on approach to Tc from above cannot be accounted for by ineffective flux penetration.
\end{thebibliography}
%

\end{document}